\documentclass[draftcls,onecolumn, 10pt]{IEEEtran}

\usepackage{silence}
\usepackage{listings}
\usepackage{signalpreamble}

\newcommand{\infectrate}{\sigma_{\text{inf}}}
\newcommand{\recoverrate}{\sigma_{\mathrm{rec}}}
\newcommand{\adit}{ADiT}
\newcommand{\setprec}{setCP-pre}
\newcommand{\setrec}{setCP-rec}
\newcommand{\setmin}{setCP-min}
\newcommand{\arbitr}{ArbiTree}
\newcommand{\signfunc}{\text{sign}}
\newcommand{\cqioc}{CQioC}

\newacronym[plural=GNNs,firstplural=Graph Neural Networks (GNNs)]{GNN}{GNN}{Graph Neural Network}
\newacronym{FDEs}{FDEs}{Fractional-order Differential Equations}
\newacronym{MSE}{MSE}{Mean Squared Error}
\newacronym{PDEs}{PDEs}{Partial Differential Equations}
\newacronym{CP}{CP}{conformal prediction}
\newacronym{CRC}{CRC}{conformal risk control}

\begin{document}

\title{Conformal Prediction for Multi-Source Detection on a Network}
\author{Xingchao Jian, Purui Zhang, Tian Lan, Feng Ji, Wenfei Liang, Wee Peng Tay, Bihan Wen, Felix Krahmer%
\thanks{
}%
}



\maketitle \thispagestyle{empty}


\begin{abstract}
    Detecting the origin of information or infection spread in networks is a fundamental challenge with applications in misinformation tracking, epidemiology, and beyond. We study the multi-source detection problem: given snapshot observations of node infection status on a graph, estimate the set of source nodes that initiated the propagation. Existing methods either lack statistical guarantees or are limited to specific diffusion models and assumptions. We propose a novel conformal prediction framework that provides statistically valid recall guarantees for source set detection, independent of the underlying diffusion process or data distribution. Our approach introduces principled score functions to quantify the alignment between predicted probabilities and true sources, and leverages a calibration set to construct prediction sets with user-specified recall and coverage levels. The method is applicable to both single- and multi-source scenarios, supports general network diffusion dynamics, and is computationally efficient for large graphs. Empirical results demonstrate that our method achieves rigorous coverage with competitive accuracy, outperforming existing baselines in both reliability and scalability. The code is available online.\footnote{\url{https://github.com/xcjian/Conformalized-Network-Source-Detection}}
\end{abstract}

\section{Introduction}\label{sect:intro}

Online social networks have experienced substantial expansion over the past few decades, becoming vital platforms for information dissemination \cite{Bak11, Mit13}. Their dense connectivity and increasing role in news distribution allow rumors, often started by multiple users, to spread quickly across large network segments. When misinformation causes harm, identifying its original sources is crucial for investigators, akin to tracing patient zero in epidemiological challenges like COVID-19 \cite{Hu20} to inform containment strategies. This source detection problem \cite{Sha11, Luo13} is the focus of our study.

The propagation of rumors or diseases on a network can be modeled as the spread of \emph{information}, where any node that acquires the information is considered \emph{infected}. The process typically begins with a finite set of source nodes, denoted by $\calY$. Each infected node may transmit the information to its neighbors and, depending on the model, may recover (e.g., in rumor spreading, a recovered node stops propagating the information) and become susceptible again. There are multiple models such as SIS, SIR, SIRS that describe different propagation patterns \cite{SheAtt:J19}. Given one or more snapshots of node statuses after the outbreak, the \emph{multi-source detection problem} aims to estimate the original set of source nodes $\calY$.

The multi-source detection problem is inherently challenging because different source sets can produce identical observed infection patterns, and exhaustively enumerating all possible source sets is computationally infeasible \cite{Sha11}. Early research primarily explored centrality-based methods \cite{Sha11, Che14, Zhu14, Wan15, Luo17, Tan18}, which estimate the sources by maximizing various graph centrality measures as proxies for the maximum likelihood estimator, relying on \emph{single snapshots} and without assuming available simulated data. However, these approaches offer provable guarantees only for specific graph structures, such as trees \cite{Sha11, Luo13}, and require heuristic adaptations for general graphs \cite{Sha11, Luo17, Ji19}. 

More recently, data-driven machine learning techniques, particularly those based on \glspl{GNN}, have been applied to multi-source detection by framing it as a node status pattern recognition problem \cite{Sha20,YanFangHe:C23}, with some leveraging \emph{multiple snapshots} \cite{ShaHasMoh:C21} and requiring knowledge about propagation model and parameter ranges for simulating training data. Despite their empirical success, these methods also lack statistical performance guarantees. 

In this work, we address this gap by developing a \gls{CP} framework \cite{Ang23,AngBarbat:25} combined with a 
backbone \gls{GNN} for multi-source detection based on multiple snapshots of node status. Unlike previous similar work using the data-driven approach, ours provides rigorous performance guarantees. Specifically, our goal is to construct a prediction set $\widehat{C}$ of nodes such that its recall rate for identifying the true source nodes exceeds $1-\beta$ with probability at least $1-\alpha$, where $\alpha, \beta \in [0,1]$ are user-specified nominal levels. The adjustable $\beta$ allows users to balance coverage and prediction set size. A smaller $\beta$ improves coverage, but also increases the size of detected node set, raising the subsequent investigation costs. 

To obtain a pre-trained \gls{GNN} and a calibration dataset for \gls{CP}, we assume access to a diffusion path dataset. This dataset can be derived from real-world diffusion examples, such as intranet network propagation, or generated through simulations when the propagation models and parameter ranges are known \cite{ShaHasMoh:C21}. While this additional information is considered expert knowledge and may require external input, it enables more accurate modeling compared to centrality-based methods. For compatibility with the SD-STGCN model \cite{ShaHasMoh:C21}, we assume multiple snapshots are available as inputs. However, our \gls{CP} approach is flexible and can accommodate various input formats.

\Gls{CP} is a statistical approach that provides confident prediction sets with guaranteed coverage, using any predictor as backbone \cite{VovGamSha:05}. It is model-free and distribution-free, ensuring the statistical coverage guarantee regardless of the predictor or data distribution. \gls{CP} is widely applied in image classification \cite{BhaWanXio:C23,AngBarbat:C24,RomSesCan:C20}, text classification \cite{TyaGuo:C23}, time series prediction \cite{BhaWanXio:C23,GibCan:J24,AngBarbat:C24}, and other inference tasks. 

In \emph{single-source detection}, traditional \gls{CP} yields a node set that contains the true source node with statistical guarantee. However, in \emph{multi-source detection} where there is a \emph{set} of source nodes, the traditional \gls{CP} methods produces a collection of node sets that contains the target node set with statistical guarantees \cite{MalPaiLen:J22,TyaGuo:C23,KatPap:J24}. This deviates from the target of finding a node set to \emph{include} a proportion ($1-\beta$) of the true source set with statistical guarantee. Existing methods like {\arbitr} and PGM \cite{CauGupDuc:J21} addressed the cases when $\beta=0$, however, they are restricted to the special case where the graph is a tree. {\adit} \cite{DawLiXu:C21} established coverage guarantee for network source detection using a hypothesis-testing approach, applicable only to the single-source case.

Another related line of research studies \gls{CP} methods for semi-supervised node classification on graphs \cite{HuaJinCan:C23,ZarSimAle:C23,ZarBoj:C24}. However, these approaches do not address the set inclusion objective required for source detection. First, while multi-source detection can be cast as a binary node classification problem, it is not a semi-supervised setting. Second, these methods provide only marginal (node-wise) coverage guarantees, rather than guarantees on the inclusion of the source set.

To the best of our knowledge, this is the first work to apply \gls{CP} to the network source detection problem using multiple snapshots and with the goal of predicting a multi-source set while providing \emph{rigorous performance guarantees}. We establish sufficient conditions for the design of non-conformity scores and the construction of prediction sets to achieve a user-specified nominal recall rate. Our approach is computationally efficient, {scaling well} with the graph size, and does not rely on restrictive assumptions about the underlying network structure. Importantly, the statistical inclusion guarantee of our method holds under the minimal assumption of exchangeability between the calibration and test data, in line with standard \gls{CP} methodology.

In summary, the main contributions of this work are as follows:
\begin{enumerate}
    \item We propose {non-conformity score designs} for set estimation problems, targeting a nominal level of recall rate with statistical guarantee. Compared to the existing methods, {these scores} yield efficient (i.e., small) prediction sets.
    \item We apply the proposed \gls{CP} approach to the multi-source detection problem. Compared to existing methods that only deal with single-source detection, this approach works under arbitrary number of sources and propagation models.
    \item We numerically validate the advantage of our proposed method on the multi-source detection problem under real-world network propagation. 
\end{enumerate}
\emph{Notation:} We use $\mathrm{Q}(\calU, a)$ to denote the sample lower $a$-quantile of a set $\calU$. The power set of a set $\calU$ is written as $2^\calU$. For a positive integer $k$, we write $[k]$ to denote the set $\set{1,2,\dots,k}$.

\section{Related Work}
\subsection{\gls{CP} for Set Estimation}
In this subsection, we present a brief review of the literature on \gls{CP} for set estimation problems. These works address the multi-label classification problem, where the target output is a set of labels, similar to the multi-source detection problem.

In multi-label classification, where the label set contains $N$ elements, any label subset can be represented as an $N$-dimensional binary vector. This allows the set estimation problem to be reformulated as a point estimation problem. However, since there are $2^N$ possible binary vectors, directly applying traditional \gls{CP} becomes computationally infeasible \cite{MalPaiLen:J22,TyaGuo:C23}. To address this, some works organize the labels into a hierarchical tree structure and apply multiple hypothesis testing to control the family-wise error rate \cite{TyaGuo:C23}. Additionally, incorporating the covariance structure between labels using the Mahalanobis distance, rather than the standard Euclidean norm, has been shown to improve prediction efficiency \cite{KatPap:J24}.

In all the aforementioned works that address \gls{CP} for label set estimation, the focus is on finding a collection of subsets that \emph{contains} the ground truth label subset with probability guarantee. This approach may not always be appropriate. In particular, for set estimation tasks such as network multi-source detection, returning a collection of possible node subsets is often impractical or uninformative. Instead, it is more desirable to provide a single node set that \emph{includes} the true source set $\calY$, or achieves a specified recall rate with high probability.

Existing methods such as PGM and {\arbitr} \cite{CauGupDuc:J21} address set inclusion problems by constructing inner and outer sets, $\widehat{C}_{\text{in}}$ and $\widehat{C}_{\text{out}}$, that satisfy $\widehat{C}_{\text{in}}\subset\calY\subset\widehat{C}_{\text{out}}$ with statistical guarantees. These approaches rely on learning a hierarchical tree structure over the label set and designing non-conformity scores tailored to this structure. The prediction sets are then efficiently computed via message passing on the label tree. However, these methods have notable limitations: (i) the tree structure learning in PGM does not scale to large label sets, as it requires solving $O(N^2)$ convex optimization problems without closed-form solutions; and (ii) both the non-conformity score design and prediction set construction are restricted to tree-structured label relationships, making them unsuitable for more general or arbitrary network structures. {\Gls{CRC} \cite{AnaSteAda:C24} is a systematic conformal prediction approach for controlling the risk {for general} set estimation. Our approach can be seen as its score-based version targeting at controlling the recall rate but with different score designs. In this work we provide a self-contained alternative proof technique to provide additional theoretical intuition for this framework.}


\subsection{Hypothesis Testing-based Confident Source Detection}

We next briefly review the confident network source detection method {\adit} \cite{DawLiXu:C21}. This method addresses the single-source detection problem over general networks under the SI model, providing a prediction set that contains the true source with statistical guarantees. For each node, a hypothesis test is formulated: the null hypothesis is that the node is the source, and the alternative is that it is not. To construct the test statistic, multiple infection paths are simulated from each node, and the average fitness of these paths with respect to the observed infection status is computed. A higher fitness indicates that the node is more likely to be the true source. Based on this fitness measure, hypothesis tests are performed for all nodes, and the prediction set is formed by including those nodes for which the null hypothesis is not rejected.

In summary, {\adit} provides assumption-free prediction sets for the source detection problem with statistical guarantees. Its fitness measure, which incorporates infection order, often leads to reasonably small prediction sets. However, the method has notable limitations: (i) it cannot be extended to multi-source scenarios, as hypothesis testing for all possible node subsets is computationally infeasible; and (ii) it cannot be directly applied to more complex models such as SIR, since the fitness function does not account for recovered nodes.

\section{Preliminaries}

\subsection{The Source Detection Problem}

In this subsection, we introduce the multi-source detection problem over networks, using a compartmental epidemic model to describe information spreading. The network is represented by a graph $G = (\calV, \calE)$, where $\calV$ is the set of nodes and $\calE$ is the set of edges, with $|\calV| = N$. 
While our approach is not restricted to a particular diffusion model, \emph{for illustration purposes}, we focus on the SIR model, which is widely used in both epidemiology and information diffusion studies \cite{ShaHasMoh:C21,Sha20}.
In this propagation model, each node belongs to one of three compartments: susceptible ($S$), infected ($I$), or recovered ($R$). A susceptible node can become infected through contact (i.e., an edge) with an infected neighbor, with a certain probability. Infected nodes may recover and transition to the recovered state.

Let $\theta_v(t) \in \set{S, I, R}$ denote the status of node $v$ at time $t$. The discrete-time dynamics are given by \cite{Kee11,Sha20,YanFangHe:C23}:
\begin{align*}
    \P(\theta_v(t) = I \mid \theta_v(t-1) = S) &= 1 - \prod_{\substack{u \in \calN(v),\\ \theta_u(t-1) = I}} (1 - \infectrate), \\
    \P(\theta_v(t) = R \mid \theta_v(t-1) = I) &= \recoverrate,
\end{align*}
where $\calN(v)$ denotes the neighbors of $v$, and $\infectrate$ and $\recoverrate$ are the infection and recovery rates, respectively. The \emph{basic reproduction number} is defined as $R_0 := \infectrate \lambda_1 / \recoverrate$, where $\lambda_1$ is the largest eigenvalue of the adjacency matrix of $G$.

The SIR model can be implemented as an independent cascade process: starting from a set of source (infected) nodes $\calY \subset \calV$, each infected node attempts to infect its neighbors independently at each time step. The SIR model generalizes the widely used SI model, which is recovered by setting $\recoverrate = 0$ (i.e., no recovery).

The \emph{multi-source detection problem} is the inverse problem that estimates the \emph{set of infection sources}, assuming snapshots of node status are observed at time points $t_1<t_2<\dots<t_M$. For any prediction set $\widehat{C}$, its performance can be measured by $\text{precision} := \frac{\abs{\widehat{C}\bigcap\calY}}{\abs{\widehat{C}}}$ and $\text{recall} := \frac{\abs{\widehat{C}\bigcap\calY}}{\abs{\calY}}$.

\subsection{Conformal Prediction}\label{subsect:preCP}

We briefly review the basic pipeline and theoretical results of conformal prediction (\gls{CP}). There are two main variants: full \gls{CP} and split \gls{CP} \cite{AngBarbat:25}. Split \gls{CP} is a computationally efficient special case of full \gls{CP} and is widely used in practice. In this work, we focus on split \gls{CP}.

Split \gls{CP} consists of the following components: (i) A \emph{point estimator}, such as a pre-trained neural network, which maps inputs to outputs. Let $\calX$ and $\calW$ denote the input and output spaces, respectively. In classification tasks, $\calW$ is the set of possible labels, and the neural network typically outputs estimated probabilities for each label. (ii) A \emph{non-conformity score} $s:\calX\times\calW\to \Real$, which quantifies how atypical a given input-output pair is with respect to the model; higher values indicate less conformity. (iii) A \emph{calibration dataset} $\set{(\bX_i, w_i)\given i\in[n]}\subset \calX\times\calW$. By evaluating the non-conformity scores on this set, we can empirically estimate the distribution of prediction errors and thus construct a prediction set. Given a new test input $\bX_{n+1}$ and a user-specified nominal level $\alpha$, split \gls{CP} defines the prediction set as $\widehat{C}(\bX_{n+1}) := \set*{w'\in\calW\given s(\bX_{n+1}, w')\leq \widehat{q}_\alpha}$,
where $\widehat{q}_\alpha := \mathrm{Q}(\set{s(\bX_{i}, w_i)}_{i=1}^n, (1-\alpha)(1+\ofrac{n}))$ is the empirical $(1-\alpha)$ quantile of the calibration scores. This construction yields the following statistical guarantee.

\begin{Theorem}[\cite{VovGamSha:05}]
If the calibration set and the test data are exchangeable, then
\begin{align}\label{eq:cover_gua}
\P(w_{n+1}\in\widehat{C}(\bX_{n+1})) \geq 1-\alpha.
\end{align}
\end{Theorem}
Among all methods satisfying the \emph{coverage} guarantee \cref{eq:cover_gua}, a smaller prediction set $\widehat{C}(\bX_{n+1})$ is preferred, as its size reflects the \emph{efficiency} of the \gls{CP} procedure.

\subsection{Conformalized Multi-source Detection}\label{subsect:probform}

We now formulate the \emph{conformalized network multi-source detection} problem. Recall that we have $N$ nodes in a graph $G$ and we observe $M$ snapshots of the node status at $M$ time instances. The input space is $\calX = \set{S,I,R}^{N\times M}$, representing the status of each node at each observed time instance. The output space is $2^\calV$, i.e., the set of all possible subsets of nodes. Suppose we have a pre-defined function $f(\bX) = (\widehat{\pi}(v))_{v\in\calV}\in\Real^N$,
where a larger $\widehat{\pi}(v)$ indicates a higher likelihood of  $v\in\calY$ given $\bX\in\calX$. For instance, $f$ can be a pre-trained neural network, and $\widehat{\pi}(v)$ can be an estimate of $\P(v\in\calY\given \bX)$. As in standard \gls{CP}, we assume access to a calibration set $\set{(\bX_i,\calY_i)\given i\in[n]}\subset \calX\times 2^\calV$, and denote the test data point as $(\bX_{n+1},\calY_{n+1})$. Given the observed snapshots $\bX_{n+1}\in\calX$, our goal is to construct a prediction set $\widehat{C}(\bX_{n+1};\alpha, \beta)$ such that
\begin{align}\label{eq:part_cov_res}
    \P( \frac{\abs*{\calY_{n+1}\cap\widehat{C}(\bX_{n+1};\alpha, \beta)}}{\abs{\calY_{n+1}}}\geq 1-\beta )\geq 1-\alpha,
\end{align}
where $\alpha$ and $\beta$ are user-specified nominal levels. In other words, the prediction set should achieve a recall rate of at least $1-\beta$ with probability at least $1-\alpha$.

\section{Conformal Prediction for Set Estimation}

In this section, we introduce a principled framework for constructing prediction sets that satisfy the recall guarantee in \cref{eq:part_cov_res}, while also achieving high efficiency in both prediction accuracy and computational cost.

\subsection{Conformal Prediction for Set Inclusion}

We first develop \gls{CP} approaches for the special case where $\beta = 0$, and then extend this to the general case where $\beta\in[0,1]$. Note that when $\beta = 0$, \cref{eq:part_cov_res} becomes
\begin{align}\label{eq:inclu_res}
    \P(\calY_{n+1}\subset \widehat{C}(\bX_{n+1};\alpha, 0)) \geq 1-\alpha.
\end{align}
In what follows, we write $\widehat{C}(\bX_{n+1};\alpha, 0)$ as $\widehat{C}(\bX_{n+1};\alpha)$ for simplicity.

\begin{figure}[!htb]
\centering
\includegraphics[scale=0.90]{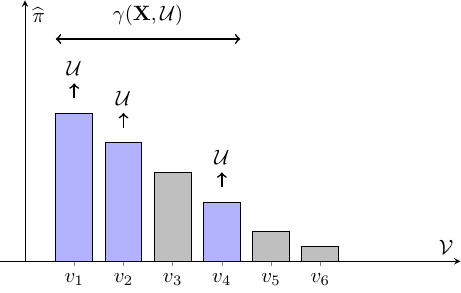}
\caption{Illustration of $\gamma(\bX,\calU)$. Suppose the nodes are indexed such that $\widehat{\pi}(v_1)\geq...\geq\widehat{\pi}(v_N)$. When $\calU = \set{v_1,v_2,v_4}$, $\gamma(\bX,\calU):=\set{v_1,v_2,v_3,v_4}$.}
\label{fig:illus_gamma}
\end{figure}

We define the following map (see \cref{fig:illus_gamma} for illustration):
\begin{align}\label{eq:gam}
    \begin{aligned}
    \gamma:\calX \times 2^{\calV}&\to 2^{\calV} \\
    (\bX, \calU)&\mapsto \set{v\in\calV \given \widehat{\pi}(v)\geq \min_{z\in\calU}\widehat{\pi}(z)}.
    \end{aligned}
\end{align}
Recall that $\widehat{\pi}(v)$ is the output of $f(\bX)$ on the vertex $v$ for any $v\in\calV$. The intuition behind $\gamma$ is that, adding those nodes from $\calV$ with larger predicted probabilities than $\min_{z\in\calU}\widehat{\pi}(z)$ to $\calU$ should yield better conformity and thus worth considering.

Let the non-conformity score be a function 
\begin{align}\label{eq:gam_based_score}
\begin{aligned}
    s:\calX\times 2^{\calV}&\to \Real \\
    (\bX, \calU)&\mapsto s(\bX, \gamma(\bX,\calU)),
\end{aligned}
\end{align}
i.e., it operates on $2^{\calV}$ through the function $\gamma$. We suppose that it is monotone: if $\gamma(\bX,\calU_1)\subset \gamma(\bX,\calU_2)$, 
\begin{align}\label{eq:mono_score}
    s(\bX, \gamma(\bX,\calU_1))\leq s(\bX, \gamma(\bX,\calU_2)).
\end{align}
Below we provide {three} examples:
{
\begin{align}
    s_{\mathrm{pre}}:
    (\bX, \calU) &\mapsto-\ofrac{\abs{\gamma(\bX,\calU)}} \sum_{z\in\gamma(\bX,\calU)} \widehat{\pi}(z),  \label{eq:prec_score} \\
    s_{\mathrm{rec}}:
    (\bX, \calU) &\mapsto\frac{\sum_{z\in\gamma(\bX,\calU)} \widehat{\pi}(z)}{\sum_{z\in\calV} \widehat{\pi}(z)}. \label{eq:recall_score} \\
    s_{\mathrm{min}}:
    (\bX, \calU) &\mapsto -\min_{z\in\gamma(\bX,\calU)} \widehat{\pi}(z). \label{eq:min_score} 
\end{align}
}
It can be shown that {$s_{\mathrm{pre}}$, $s_{\mathrm{rec}}$ and $s_{\mathrm{min}}$} satisfy \cref{eq:gam_based_score,eq:mono_score}.

\begin{Remark}
The score designs \cref{eq:prec_score,eq:recall_score} are meaningful since they represent the non-conformity in terms of precision and recall rate respectively, under the oracle case where $\widehat{\pi}(v) = \P(v\in\calY_{n+1}\given \bX_{n+1})$ for all $v\in\calV$.
For any fixed $\calU\subset\calV$, the expected precision of the set $\gamma(\bX,\calU)$ given $\bX_{n+1}$ can be calculated as
\begin{align*}
    & \E[\frac{\abs{\gamma(\bX,\calU)\bigcap\calY_{n+1}}}{\abs{\gamma(\bX,\calU)}}\given \bX_{n+1}] \\
    &= \ofrac{\abs{\gamma(\bX,\calU)}}\sum_{z\in\gamma(\bX,\calU)} \E[\indicate{z\in\calY_{n+1}}\given \bX_{n+1}],
\end{align*}
which coincides with \cref{eq:prec_score} under the oracle case up to a minus sign.
Using similar arguments, the recall rate of the set $\gamma(\bX,\calU)$ given $\bX_{n+1}$ can be calculated as
\begin{align*}
    \frac{\E[\abs{\gamma(\bX,\calU)\bigcap\calY_{n+1}}\given \bX_{n+1}]}{\E[\abs{\calY_{n+1}}\given \bX_{n+1}]} = \frac{\sum_{z\in\gamma(\bX,\calU)} \widehat{\pi}(z)}{\sum_{z\in\calV} \widehat{\pi}(z)},
\end{align*}
which coincides with \cref{eq:recall_score}.

In \cref{sect:comp_crc} we show that The score design \cref{eq:min_score} yields the same prediction set with the one proposed by \cite{AnaSteAda:C24}.
\end{Remark}

Let $\widehat{q}_\alpha:= \mathrm{Q}(\set{s(\bX_i, \calY_i)}_{i=1}^n, (1-\alpha)(1+\ofrac{n}))$. The prediction set is defined as 
\begin{align}\label{eq:confmset}
\widehat{C}(\bX_{n+1};\alpha) := \set*{v\in\calV\given s(\bX_{n+1}, \set{v})\leq \widehat{q}_\alpha}.
\end{align}
Next, we prove that $\widehat{C}(\bX_{n+1};\alpha)$ satisfies \cref{eq:inclu_res}. This is the result of the following key observation.

\begin{Proposition}\label{prop:equi_cover}
Under the conditions \cref{eq:gam_based_score} and  \cref{eq:mono_score}, the following two events are equivalent:
\begin{align}
\calY_{n+1}&\subset\widehat{C}(\bX_{n+1};\alpha); \label{eq:event_subset}\\
s(\bX_{n+1}, \calY_{n+1})&\leq \widehat{q}_\alpha.\label{eq:event_leq}
\end{align}
\end{Proposition}
\begin{proof}
    We first prove that \cref{eq:event_subset} implies \cref{eq:event_leq}. Let $z_0:=\argmin_{z\in\calY_{n+1}}\widehat{\pi}(z)$. Note that $s(\bX_{n+1}, \calY_{n+1}) = s(\bX_{n+1}, \set{z_0})$, hence $z_0\in\calY_{n+1}\subset\widehat{C}(\bX_{n+1};\alpha)$. Hence $s(\bX_{n+1}, \calY_{n+1}) = s(\bX_{n+1}, \set{z_0})\leq \widehat{q}_\alpha$.

    Next we prove that \cref{eq:event_leq} implies \cref{eq:event_subset}. Note that $s(\bX_{n+1}, \set{z_0}) = s(\bX_{n+1}, \calY_{n+1})\leq \widehat{q}_\alpha$. For any $z'\in\calY_{n+1}$, we know that $\gamma(\bX,\set{z'})\subset \gamma(\bX,\set{z_0})$, hence $s(\bX_{n+1}, \set{z'})\leq s(\bX_{n+1}, \set{z_0})\leq \widehat{q}_\alpha$ by \cref{eq:mono_score}. Therefore $z'\in\widehat{C}(\bX_{n+1};\alpha)$, which completes the proof.
\end{proof}

\Cref{prop:equi_cover} yields the inclusion guarantee \cref{eq:inclu_res} as follows.

\begin{Theorem}\label{thm:inclu_res}
    Assume that $\set{(\bX_i,\calY_i)\given i\in[n+1]}$ are exchangeable. Then $\widehat{C}(\bX_{n+1};\alpha)$ (cf. \ \cref{eq:confmset}) satisfies the inclusion guarantee \cref{eq:inclu_res}.
\end{Theorem}
\begin{proof}
    Since $\set{(\bX_i,\calY_i)\given i\in[n+1]}$ are exchangeable, $\set{s(\bX_i,\calY_i)\given i\in[n+1]}$ are also exchangeable. According to \cite[Fact 2.14 (ii) and Lemma 3.4]{AngBarbat:25}, we know that $\P(s(\bX_{n+1}, \calY_{n+1})\leq \widehat{q}_\alpha) \geq 1-\alpha$. Combining this with \cref{prop:equi_cover} we know that \cref{eq:inclu_res} holds true.
\end{proof}

Furthermore, in \cref{subsect:diffconn,sect:comp_crc}, we demonstrate that \cref{eq:confmset} can also be obtained by applying our proposed scores within existing \gls{CP} frameworks.

\subsection{Conformal Prediction with Recall Rate Guarantee}

\Cref{thm:inclu_res} establishes the validity of the prediction set construction in \cref{eq:confmset} for the special case where the nominal recall rate is $1$ (i.e., $\beta=0$ in \cref{eq:part_cov_res}). We now generalize this approach to handle arbitrary recall levels $\beta \in [0,1]$. Our strategy consists of two steps: first, we define a map that shrinks the ground truth set $\calY$ by retaining only a $(1-\beta)$ fraction of its elements; second, we apply the prediction set construction from \cref{eq:confmset} to ensure inclusion of this shrunken set.

To shrink sets, we define the map $\nu: \calX\times 2^\calV \to 2^\calV$ such that 
\begin{align}
    \nu(\bX, \calU)&\subset \calU \label{eq:maptosubset}\\ 
    \abs{\nu(\bX, \calU)}&\geq (1-\beta)\abs{\calU}. \label{eq:shrinkset}
\end{align}
for any $\calU\subset\calV$. Then replacing $\calY_i$ by the shrunken sets $\nu(\bX_i, \calY_i)$ for all $i\in[n+1]$ and formulating the prediction set via \cref{eq:confmset}, we obtain a set $\widehat{C}(\bX_{n+1};\alpha, \beta)$ such that
\begin{align}\label{eq:cover_subset}
    \P(\nu(\bX_{n+1}, \calY_{n+1})\subset\widehat{C}(\bX_{n+1};\alpha, \beta))\geq 1-\alpha.
\end{align}
Note that $\nu(\bX_{n+1}, \calY_{n+1})\subset\widehat{C}(\bX_{n+1};\alpha, \beta)$ implies
\begin{align*}
    \frac{\abs*{\calY_{n+1}\bigcap\widehat{C}(\bX_{n+1};\alpha, \beta)}}{\abs{\calY_{n+1}}}\geq 1-\beta.
\end{align*}
Combining this with \cref{eq:cover_subset} we obtain the following result.

\begin{Theorem}\label{thm:inclu_res_part}
Suppose the map $\nu$ satisfies \cref{eq:maptosubset,eq:shrinkset}. Let 
\begin{align*}
\widehat{q}_\alpha:= \mathrm{Q}\parens*{ \set*{s(\bX_i, \nu(\bX_i, \calY_i))}_{i=1}^n, (1-\alpha)(1+\ofrac{n}) },
\end{align*}
and define the prediction set as
\begin{align}\label{eq:confmset_part}
\widehat{C}(\bX_{n+1};\alpha,\beta) := \set*{v\in\calV\given s(\bX_{n+1}, \set{v})\leq \widehat{q}_\alpha}.
\end{align}
Then, this set satisfies the recall rate guarantee \cref{eq:part_cov_res}.
\end{Theorem}
\begin{proof}
    Define $\widetilde{\calY}_i:=\nu(\bX_i,\calY_i )$ for $i\in[n+1]$. By applying \cref{thm:inclu_res} on the pairs $\set{(\bX_i, \widetilde{\calY}_i)}$, we know that
    \begin{align*}
        \P(\widetilde{\calY}_i\subset\widehat{C}(\bX_{n+1};\alpha,\beta))\geq 1-\alpha. 
    \end{align*}
    On the other hand, since $\widetilde{\calY}_i\subset\widehat{C}(\bX_{n+1};\alpha,\beta)$ implies $\frac{\abs{\calY_i\bigcap \widehat{C}(\bX_{n+1};\alpha,\beta)}}{\abs{\calY_i}}\geq 1-\beta$, we have
    \begin{align*}
        \P(\frac{\abs{\calY_i\bigcap \widehat{C}(\bX_{n+1};\alpha,\beta)}}{\abs{\calY_i}}\geq 1-\beta)\geq 1-\alpha. 
    \end{align*}
    This completes the proof.
\end{proof}

\begin{figure}[!htb]
\centering
\includegraphics[scale=0.90]{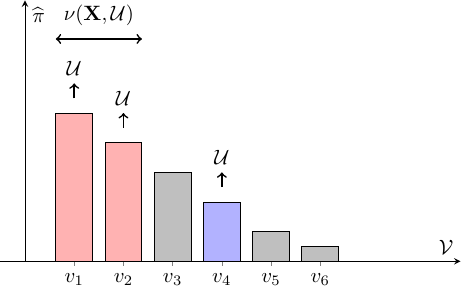}
\caption{Illustration of $\nu(\bX,\calU)$. Suppose the nodes are indexed such that $\widehat{\pi}(v_1)\geq...\geq\widehat{\pi}(v_N)$. When $\calU = \set{v_1,v_2,v_4}$, $\beta=\ofrac{3}$, $\nu(\bX,\calU)=\set{v_1,v_2}$.}
\label{fig:illus_nu}
\end{figure}

In the above argument, the only requirements on the shrinking map $\nu$ are \cref{eq:maptosubset,eq:shrinkset}, i.e., $\nu(\bX, \calU)$ should be a subset of $\calU$ while maintaining at least $(1-\beta)$ proportion of elements. One natural construction of such a map is to retain the $(1-\beta)$ proportion of elements with the largest predicted probability (cf.\ \cref{fig:illus_nu}), i.e.,
\begin{align}\label{eq:ordered_shrink}
\begin{aligned}
    &\nu(\bX_{n+1}, \calY_{n+1}) \\
    &:= \set*{v\in\calY_{n+1}\given \widehat{\pi}(v)\geq -\mathrm{Q}(\set{-\widehat{\pi}(v)}_{v\in\calY_{n+1}}, 1-\beta)}.
\end{aligned}
\end{align}

Finally, we note that this shrinking strategy can be combined with any existing \gls{CP} approach that achieves the inclusion guarantee in \cref{eq:inclu_res}, such as {\arbitr} \cite{CauGupDuc:J21}. By applying the same reasoning, the resulting method also satisfies the recall rate guarantee in \cref{eq:part_cov_res}. {It can be shown that when using $s_\mathrm{min}$ in \cref{eq:confmset_part}, the resulting conformal set coincides with that of \cite[(7)]{AnaSteAda:C24} under the quantile risk control scenario.} 

\section{Experiments}

\begin{table*}[htbp]
    \centering
    \small 
    \caption{Inclusion rates and prediction set sizes under the SIR model with random parameters ($|\mathcal{Y}| \in [15]$, $R_0 \in [1, 15]$, $\recoverrate \in [0.1, 0.4]$) over the highSchool network ($N=774$) for methods \setprec, \setrec, and \arbitr. The best and the second-best results under each $(\alpha,\beta)$ configuration are {\bf boldfaced} and \underline{underlined}, respectively.} 
    \begin{tabular}{ll *{6}{c@{\hspace{0.3cm}}}}
        \toprule
        & & \multicolumn{3}{c}{\textbf{Inclusion Rates}} & \multicolumn{3}{c}{\textbf{Prediction Set Size}} \\
        \cmidrule(lr){3-5} \cmidrule(lr){6-8}
        \textbf{Method} & \textbf{$\beta$} & \textbf{$\alpha=0.05$} & \textbf{$\alpha=0.10$} & \textbf{$\alpha=0.15$} & \textbf{$\alpha=0.05$} & \textbf{$\alpha=0.10$} & \textbf{$\alpha=0.15$} \\
        \midrule
        \multirow{4}{*}{\textbf{\setrec}} & $0.1$ & 1.000 $\pm$ 0.000 & 1.000 $\pm$ 0.000 & 1.000 $\pm$ 0.000 & 774.000 $\pm$ 0.000 & 774.000 $\pm$ 0.000 & 774.000 $\pm$ 0.000\\
        & $0.3$ & 1.000 $\pm$ 0.000 & 0.897 $\pm$ 0.012 & 0.845 $\pm$ 0.016 & 774.000 $\pm$ 0.000 & \underline{16.246 $\pm$ 0.479} & \underline{15.017 $\pm$ 0.434} \\
        & $0.5$ & 0.950 $\pm$ 0.010 & 0.899 $\pm$ 0.013 & 0.847 $\pm$ 0.015 & \textbf{11.856 $\pm$ 0.338} & \textbf{9.874 $\pm$ 0.279} & \textbf{8.912 $\pm$ 0.248} \\
        & $0.7$ & 0.947 $\pm$ 0.010 & 0.898 $\pm$ 0.014 & 0.848 $\pm$ 0.016 & \textbf{8.949 $\pm$ 0.273} & \textbf{6.570 $\pm$ 0.186} & \textbf{5.307 $\pm$ 0.148} \\
        \midrule
        \multirow{4}{*}{\textbf{\setprec}} & $0.1$ & 0.951 $\pm$ 0.011 & 0.899 $\pm$ 0.016 & 0.852 $\pm$ 0.016 & \underline{25.069 $\pm$ 0.566} & \underline{22.576 $\pm$ 0.511} & \underline{20.640 $\pm$ 0.463} \\
        & $0.3$ & 0.951 $\pm$ 0.012 & 0.899 $\pm$ 0.016 & 0.850 $\pm$ 0.018 & \underline{21.333 $\pm$ 0.486} & 18.622 $\pm$ 0.429 & 16.599 $\pm$ 0.391\\
        & $0.5$ & 0.949 $\pm$ 0.009 & 0.898 $\pm$ 0.015 & 0.847 $\pm$ 0.017 & 16.493 $\pm$ 0.384 & 13.670 $\pm$ 0.327 & 11.719 $\pm$ 0.291\\
        & $0.7$ & 0.948 $\pm$ 0.009 & 0.899 $\pm$ 0.014 & 0.849 $\pm$ 0.016 & 13.056 $\pm$ 0.307 & 9.669 $\pm$ 0.283 & 7.860 $\pm$ 0.273\\
        \midrule
        \multirow{4}{*}{\textbf{\setmin}} & $0.1$ &  0.949 ± 0.011  &  0.900 ± 0.015  &  0.852 ± 0.018  & \textbf{19.401 ± 0.517} & \textbf{18.503 ± 0.476} & \textbf{17.469 ± 0.447} \\
        & $0.3$ &  0.953 ± 0.012  &  0.901 ± 0.015  &  0.850 ± 0.017  & \textbf{17.165 ± 0.441} & \textbf{15.649 ± 0.416} & \textbf{14.484 ± 0.395} \\
        & $0.5$ &  0.946 ± 0.010  &  0.897 ± 0.013  &  0.848 ± 0.019  & \underline{13.739 ± 0.443} & \underline{12.436 ± 0.416} & \underline{11.272 ± 0.395} \\
        & $0.7$ &  0.949 ± 0.009  &  0.897 ± 0.015  &  0.849 ± 0.017  & \underline{10.308 ± 0.286} & \underline{8.057 ± 0.254} & \underline{6.761 ± 0.255} \\
        \midrule
        \multirow{4}{*}{\textbf{\arbitr}} & $0.1$ & 0.993 $\pm$ 0.005 & 0.980 $\pm$ 0.008 & 0.961 $\pm$ 0.010 & 771.022 $\pm$ 1.054 & 764.048 $\pm$ 4.216 & 750.241 $\pm$ 10.979\\
        & $0.3$ & 0.992 $\pm$ 0.005 & 0.973 $\pm$ 0.009 & 0.941 $\pm$ 0.014 & 763.480 $\pm$ 3.888 & 741.736 $\pm$ 7.911 & 709.067 $\pm$ 10.800\\
        & $0.5$ & 0.983 $\pm$ 0.007 & 0.948 $\pm$ 0.013 & 0.910 $\pm$ 0.014 & 748.169 $\pm$ 6.251 & 710.206 $\pm$ 10.059 & 669.028 $\pm$ 11.733\\
        & $0.7$ & 0.969 $\pm$ 0.008 & 0.928 $\pm$ 0.012 & 0.881 $\pm$ 0.014 & 727.488 $\pm$ 9.241 & 681.732 $\pm$ 10.512 & 637.896 $\pm$ 12.525 \\
        \bottomrule
    \end{tabular}
    \label{tab:nomilevels}
\end{table*}

\begin{table*}[htbp]
    \centering
    \small 
    \caption{Inclusion rates and prediction set sizes under the SI model ($\infectrate=0.25$) and SIR model ($\infectrate=0.25$, $\recoverrate=0.15$) over the highSchool network ($N=774$) with different number of sources (denoted by $\abs{\calY}$). When $\abs{\calY} > 1$, nominal levels are set as $\alpha=0.1$, $\beta=0.3$. When $\abs{\calY}=1$, $\beta = 0$.}
    \begin{tabular}{ll *{6}{c@{\hspace{0.3cm}}}}
        \toprule
        & & \multicolumn{3}{c}{\textbf{Inclusion Rates}} & \multicolumn{3}{c}{\textbf{Prediction Set Size}} \\
        \cmidrule(lr){3-5} \cmidrule(lr){6-8}
        \textbf{Model} & \textbf{Method} & \textbf{$|\mathcal{Y}|=1$} & \textbf{$|\mathcal{Y}|=7$} & \textbf{$|\mathcal{Y}|=10$} & \textbf{$|\mathcal{Y}|=1$} & \textbf{$|\mathcal{Y}|=7$} & \textbf{$|\mathcal{Y}|=10$} \\
        \midrule
        \multirow{4}{*}{\textbf{SI}} 
        & \setrec & 0.901 $\pm$ 0.015 & 0.903 $\pm$ 0.016 & 0.899 $\pm$ 0.014 & \underline{23.315 $\pm$ 0.626} & \textbf{29.053 $\pm$ 0.288} & \textbf{40.309 $\pm$ 0.437} \\
        & \setprec & 0.902 $\pm$ 0.013 & 0.900 $\pm$ 0.014 & 0.904 $\pm$ 0.015 & 25.691 $\pm$ 0.324 & 37.438 $\pm$ 0.372 & 52.139 $\pm$ 0.513 \\
        & \setmin & 0.901 ± 0.015 & 0.905 ± 0.014 & 0.902 ± 0.017 & \textbf{20.442 ± 0.449} & \underline{30.216 ± 0.288} & \underline{41.507 ± 0.358} \\
        & \adit & 0.988 $\pm$ 0.005 & - & - & 24.432 $\pm$ 0.698 & - & - \\
        & \arbitr & 0.896 $\pm$ 0.014 & 1.000 $\pm$ 0.001 & 1.000 $\pm$ 0.001 & 624.442 $\pm$ 14.545 & 770.490 $\pm$ 1.087 & 770.933 $\pm$ 0.774 \\
        \midrule
        \multirow{4}{*}{\textbf{SIR}} & \textbf{Method} & $|\mathcal{Y}|\!=\!1$ & $|\mathcal{Y}|\!=\!7$ & $|\mathcal{Y}|\!=\!10$ & $|\mathcal{Y}|\!=\!1$ & $|\mathcal{Y}|\!=\!7$ & $|\mathcal{Y}|\!=\!10$ \\
        \cmidrule(lr){2-8}
        & \setrec & 0.901 $\pm$ 0.015 & 0.903 $\pm$ 0.017 & 0.903 $\pm$ 0.015 & \underline{23.045 $\pm$ 0.607} & \textbf{27.996 $\pm$ 0.308} & \textbf{37.732 $\pm$ 0.311} \\
        & \setprec & 0.900 $\pm$ 0.015 & 0.902 $\pm$ 0.017 & 0.901 $\pm$ 0.015 & 23.551 $\pm$ 0.366 & 34.451 $\pm$ 0.287 & 49.255 $\pm$ 0.502 \\
        & \setmin & 0.898 ± 0.013 & 0.901 ± 0.016 & 0.904 ± 0.013 & \textbf{19.171 ± 0.457} & \underline{28.951 ± 0.353} & \underline{41.553 ± 0.467} \\
        & \arbitr & 0.902 $\pm$ 0.016 & 1.000 $\pm$ 0.000 & 1.000 $\pm$ 0.000 & 607.128 $\pm$ 13.636 & 768.980 $\pm$ 0.863 & 769.771 $\pm$ 1.178 \\
        \bottomrule
    \end{tabular}
    
    \label{tab:si_sir}
\end{table*}

\begin{table*}[htbp]
    \centering
    \small 
    \caption{Inclusion rates and prediction set sizes under the SIR model with random parameters ($|\mathcal{Y}| \in [15]$, $\recoverrate \in [0.1, 0.4]$) over the highSchool, bkFratB, and sfhh networks ($N=774,58,403$, respectively). $R_0$ are drawn from different ranges to represent different propagation speed. Nominal levels are set as $\alpha=0.1$, $\beta=0.3$.}
    \begin{tabular}{ll *{6}{c@{\hspace{0.25cm}}}}
        \toprule
        & & \multicolumn{3}{c}{\textbf{Inclusion Rates}} & \multicolumn{3}{c}{\textbf{Prediction Set Size}} \\
        \cmidrule(lr){3-5} \cmidrule(lr){6-8}
        \textbf{Method} & \textbf{$R_0$} & \textbf{highSchool} & \textbf{bkFratB} & \textbf{sfhh} & \textbf{highSchool} & \textbf{bkFratB} & \textbf{sfhh} \\
        \midrule
        \multirow{3}{*}{\textbf{\setrec}} 
        & $[1, 15]$ & 0.897 $\pm$ 0.012 & 0.902 $\pm$ 0.016 & 0.901 $\pm$ 0.014 & \underline{16.246 $\pm$ 0.479} & \textbf{13.415 $\pm$ 0.426} & \underline{14.740 $\pm$ 0.485} \\
        & $[11, 25]$ & 0.894 $\pm$ 0.018 & 0.902 $\pm$ 0.015 & 0.902 $\pm$ 0.016 & \underline{26.340 $\pm$ 0.783} & \textbf{19.425 $\pm$ 0.569} & \underline{22.862 $\pm$ 0.859} \\
        & $[21, 35]$ & 0.897 $\pm$ 0.015 & 0.895 $\pm$ 0.017 & 0.901 $\pm$ 0.017 & \textbf{37.055 $\pm$ 1.270} & \textbf{23.763 $\pm$ 0.547} & \textbf{30.245 $\pm$ 0.965} \\
        \midrule
        \multirow{3}{*}{\textbf{\setprec}} 
        & $[1, 15]$ & 0.899 $\pm$ 0.016 & 0.896 $\pm$ 0.014 & 0.899 $\pm$ 0.013 & 18.622 $\pm$ 0.429 & 16.256 $\pm$ 0.478 & 16.107 $\pm$ 0.441 \\
        & $[11, 25]$ & 0.898 $\pm$ 0.018 & 0.900 $\pm$ 0.016 & 0.900 $\pm$ 0.015 & 31.705 $\pm$ 0.902 & 24.867 $\pm$ 0.778 & 25.854 $\pm$ 0.651 \\
        & $[21, 35]$ & 0.904 $\pm$ 0.016 & 0.895 $\pm$ 0.015 & 0.901 $\pm$ 0.013 & 47.044 $\pm$ 1.553 & 31.433 $\pm$ 0.842 & 36.330 $\pm$ 0.941 \\
        \midrule
        \multirow{3}{*}{\textbf{\setmin}} 
        & $[1, 15]$ & 0.901 ± 0.015 & 0.897 ± 0.013 & 0.900 ± 0.016 & \textbf{15.649 ± 0.416}  & \underline{14.043 ± 0.453}  &  \textbf{14.084 ± 0.426}  \\
        & $[11, 25]$ & 0.901 ± 0.015 & 0.899 ± 0.016 & 0.901 ± 0.016 & \textbf{26.091 ± 0.654} &  \underline{21.273 ± 0.657}  & \textbf{21.998 ± 0.598}  \\
        & $[21, 35]$ & 0.903 ± 0.014 & 0.898 ± 0.016 & 0.900 ± 0.013 &  \underline{37.698 ± 1.262}  &  \underline{26.472 ± 0.691} & \underline{30.313 ± 0.767}  \\
        \midrule
        \multirow{3}{*}{\textbf{\arbitr}} 
        & $[1, 15]$ & 0.973 $\pm$ 0.009 & 0.988 $\pm$ 0.010 & 0.973 $\pm$ 0.013 & 741.736 $\pm$ 7.911 & 56.618 $\pm$ 0.683 & 384.167 $\pm$ 4.975 \\
        & $[11, 25]$ & 0.992 $\pm$ 0.004 & 0.996 $\pm$ 0.005 & 0.980 $\pm$ 0.008 & 761.811 $\pm$ 3.511 & 57.528 $\pm$ 0.398 & 390.277 $\pm$ 3.583 \\
        & $[21, 35]$ & 0.990 $\pm$ 0.006 & 0.998 $\pm$ 0.004 & 0.987 $\pm$ 0.006 & 762.320 $\pm$ 4.470 & 57.816 $\pm$ 0.264 & 394.993 $\pm$ 2.315 \\
        \bottomrule
    \end{tabular}
    
    \label{tab:diffgraphs}
\end{table*}

In this section, we present numerical results for conformalized multi-source detection under various propagation patterns on real-world networks. Our objectives are twofold: (i) to achieve the recall rate guarantee in \cref{eq:part_cov_res}; and (ii) to maximize efficiency by minimizing prediction set sizes. 

We implement our method using {the} non-conformity score designs {$s_{\mathrm{pre}}$, $s_{\mathrm{rec}}$ and $s_{\mathrm{min}}$} (see {\cref{eq:prec_score,eq:recall_score,eq:min_score}}), referred to as {{\setprec}, {\setrec} and {\setmin}}, respectively. For comparison, we include the following baselines: {\adit} \cite{DawLiXu:C21} and {\arbitr} \cite{CauGupDuc:J21}. Recall that our proposed method with the non-conformity score $s_{\mathrm{min}}$ yields the same prediction set with \cite{AnaSteAda:C24}, hence can also be regarded as a baseline. All methods except {\adit} are based on the conformal prediction (\gls{CP}) framework and thus require a pre-trained neural network. We use the SD-STGCN model \cite{ShaHasMoh:C21} as the backbone, modifying its output to $N\times2$ channels and adapting the loss function for binary node classification in the multi-source detection setting.

We evaluate the proposed and baseline methods across a range of nominal levels $(\alpha, \beta)$ to assess their ability to achieve the desired recall guarantees. In addition, we systematically investigate performance under various practical scenarios, including: (i) varying the number of sources; (ii) different diffusion models (e.g., SI and SIR); (iii) a range of propagation speeds; and (iv) multiple real-world network topologies.

\textbf{Datasets.} We simulate SIR and SI propagation processes over three social networks: highSchool, bkFratB, and sfhh, which are obtained by aggregating contact records \cite{ShaHasMoh:C21} within groups of people.

\textbf{Experimental setup.}
For all experiments, we use a calibration set of size $n=7600$ and a test set of size $400$. The pre-trained model is trained on a separate set of $20,000$ samples. Each experiment is repeated over $50$ random splits of the calibration and test sets to ensure robustness of the results. All performance differences are significant in terms of a level 0.95 Wilcoxon signed-rank test. 

Further details regarding the dataset and experimental settings are included in \cref{sect:expset,sect:addexpres}.

\subsection{Conformalized Detection under Different Nominal Levels}\label{subsect:expnominals}

We evaluate the quality of prediction sets derived by different methods under different combinations of nominal levels $\alpha$ and $\beta$. According to \cref{eq:part_cov_res}, it is expected that the prediction sets cover at least $(1-\beta)$ proportion of the source set with probability at least $(1-\alpha)$. Besides, the size of the prediction sets should be as small as possible. We implement the methods under SIR models with parameters $\infectrate$, $\recoverrate$, $R_0$ randomly drawn from uniform distributions over certain ranges. Since {\adit} cannot be applied under multi-source detection problem, it is not included in this comparison.

In \cref{tab:nomilevels}, we report the empirical results for different combinations of $(\alpha, \beta)$. Across all settings, each method achieves the desired recall rate $(1-\beta)$ with probability at least $(1-\alpha)$ (up to a standard error). This provides empirical support for \cref{thm:inclu_res_part}, confirming the validity of the statistical guarantees for our proposed non-conformity scores and prediction set constructions. Furthermore, these results demonstrate that the shrinking strategy using the map $\nu$ (see \cref{eq:ordered_shrink}) effectively calibrates the recall rate to the user-specified nominal level, not only for our proposed set construction in \cref{eq:confmset_part}, but also for alternative methods such as {\arbitr}.

Across most nominal level settings, {\setrec}, {\setprec} and {\setmin} produce substantially smaller prediction sets than {\arbitr}. In contrast, {\arbitr} yields prediction sets that are nearly as large as the entire graph ($N=774$), making them impractically large for informative inference. When the nominal recall rate is high (i.e., $\beta$ is small), {\setrec} also produces large, uninformative prediction sets. This limitation arises from its score definition in \cref{eq:recall_score}. In the extreme case where $\beta=0$ (requiring all source nodes to be included), if the \gls{GNN} assigns near-zero probability to any true source node $v_0$, after being divided by the denominator $\sum_{z\in\calV} \widehat{\pi}(z)$, the value $\frac{\widehat{\pi}(v_0)}{\sum_{z\in\calV} \widehat{\pi}(z)}$ becomes almost zero. Hence $s_{\mathrm{rec}}$ will result in the entire node set $\calV$ being selected. In contrast, the score $s_{\mathrm{min}}$ is less affected as the small (near-zero) probabilities are not scaled. The score $s_{\mathrm{pre}}$, which incorporates set size into its formulation, also yields more efficient (smaller) prediction sets than $s_{\mathrm{rec}}$ in such scenarios.

Finally, we remark that, since all methods fulfill the statistical guarantees, in practice, different \gls{CP} approaches can be utilized, and the smallest prediction set can be chosen.

\subsection{Conformalized Detection under Different Propagation Models}\label{subsect:exppropmods}

We next compare the prediction performance of different methods across various propagation models and network topologies. The observed inclusion rates further confirm the statistical guarantees established in \cref{thm:inclu_res_part}, while the sizes of the prediction sets provide insight into the inherent difficulty of each scenario.

In \cref{tab:si_sir}, we present the numerical results for both SI and SIR models with varying numbers of sources. All methods achieve the desired recall rate guarantee, while {\setrec} and {\setmin} consistently yields the smallest and the second smallest prediction sets. Additionally, for a fixed number of sources and method, the prediction sets under the SIR model are always smaller than those under the SI model. This is because the SIR model provides additional information through the recovery status, making the inverse problem easier and resulting in more efficient (smaller) prediction sets.

\Cref{tab:diffgraphs} reports results across different social networks. As before, all methods satisfy the recall rate guarantee, and {\setrec} consistently achieves the smallest prediction sets. For a fixed network and method, the prediction set size increases with larger values of $R_0$. This is expected, as a higher $R_0$ corresponds to a faster infection rate $\infectrate$, making the inverse problem more challenging.

\begin{table}[!tb]
    \centering 
    \small
    \caption{Average time (in seconds) for computing on $400$ test samples under the SI model ($\infectrate=0.25$), $\abs{\calY}=1$, with $\alpha = 0.05, 0.07, 0.10, 0.15, 0.20$.}
    \begin{tabular}{lc}
        \toprule
        \textbf{Method} & \textbf{Time Cost} \\
        \midrule
        \setrec  & $1.392 \pm 0.008$ \\
        \setprec & $1.559 \pm 0.008$ \\
        \setmin  & $1.249 \pm 0.009$ \\
        \adit    & $792.170 \pm 23.936$ \\
        \arbitr  & $79.637 \pm 1.208$ \\
        \bottomrule
    \end{tabular}
    \label{tab:time_cost}
\end{table}

The execution time for evaluating each method on the test data is shown in \cref{tab:time_cost}. {\adit} and {\arbitr} are implemented by parallel computing to manage their computational demands, as using a single thread would result in excessive time costs. In contrast, {\setrec}, {\setprec} and {\setmin} operate on a single thread without parallel computing. The results demonstrate that {\setrec}, {\setprec} and {\setmin} are substantially faster than both {\adit} and {\arbitr}. Although the computational complexity for evaluating the prediction set in \cref{eq:confmset_part} is {$O(N\log N)$—slightly larger than {\arbitr} and {\adit} with complexity $O(N)$}—the practical runtimes for evaluating \cref{eq:confmset_part} is significantly smaller. This is because {\arbitr} requires two recursive message-passing steps over the label tree, whereas {\setrec} and {\setprec} only require a single pass to compute scores for all nodes followed by sorting. In contrast, {\adit} performs a hypothesis test for each node, which involves estimating the distribution of the test statistic via multiple Monte Carlo simulations, further increasing its computational cost.

\section{Conclusion and Limitations}

In this paper, we introduced a systematic \gls{CP} framework for network multi-source detection, capable of achieving user-specified recall rates with rigorous statistical guarantees. Theoretically, our approach relies only on the minimal assumption of data exchangeability, offering a general solution to conformalized set estimation problems. Empirically, the method demonstrates superior efficiency compared to existing approaches, both in terms of prediction accuracy and computational performance.

One limitation of our approach is its dependence on having access to a calibration set of diffusion data. In some applications, this can be synthetically generated based on prior knowledge of the propagation model and parameter ranges, which may require external expert input. Future work involves studying how mismatch in this prior knowledge affects the performance of our method, and how to adaptively update our prediction using new data.

\section{Comparison and Relation with Existing Approaches}\label[Appendix]{sect:comp}

In this section, we introduce the score design PGM and {\arbitr} and prediction set formulation {\cqioc} \cite{CauGupDuc:J21}, and analyze the inherent connection and advantage of our proposed method. 

\subsection{Definitions and Notations}
To facilitate presentation, we first introduce relevant notations. We define the following sign function, which converts any subset of $\calV$ to a binary vector:
\begin{align*}
    \text{sign}: 2^{\calV} &\mapsto \set{-1, 1}^{N} \\
    \calU &\to 2(\indicate{\calU} - \ofrac{2}),
\end{align*}
i.e., $\signfunc(\calU)_{v}=1$ if and only if $v\in\calU$, otherwise $\signfunc(\calU)_{v}=-1$. Therefore, each vertex $v$ is either classified as $-1$ or $1$. 

Recall that, $f$ outputs the estimate of $\P(v\in\calY\given \bX)$ for each $v$. Therefore, for each $v$ a function $\widetilde{f}_v:\calX\to\Real$ may be defined such that $\widetilde{f}_v(\bX)$ takes large value if $v$ is recognized to be more likely in $\calY$ according to $f$. For example, $\widetilde{f}_v$ could be the logistic regressor on $v$ with inputs being the output feature of $f$. Given $\widetilde{f}_v$, the following functions are defined:
\begin{align*}
    \varphi_v:\set{-1, 1}\times\calX&\mapsto \Real^2 \\
    (i,\bX)&\to \ofrac{2}\begin{pmatrix}
        (i-1)\cdot \widetilde{f}_v(\bX)\\
        (i+1)\cdot \widetilde{f}_v(\bX)
    \end{pmatrix}, \\
    \psi:\set{-1, 1}^2&\mapsto \Real^4,
\end{align*}
where $\psi$ is a bijection between $\set{-1, 1}^2$ and the standard basis in $\Real^4$.

\subsection{Non-conformity Score and Prediction Set Design}
To incorporate the relation between nodes and facilitate computation, it is assumed that $G$ has a tree structure \cite{CauGupDuc:J21}. Depending on the output type of $f$, the following scores were designed:\footnote{To be consistent with the main content where a smaller score indicates better conformity, here we have added a negative sign based on their formulation in the original paper \cite{CauGupDuc:J21}.}
\begin{align}\label{eq:arbitrscore}
    \begin{aligned}   
    s_{\text{ArbiTree}}(\bX, \calU) &= - \Big(\sum_{v\in\calV} c_{v}^{\text{ArbiTree}}\signfunc(\calU)_{v}\signfunc(f(\bX))_{v} \\
    &+ \sum_{e = (v_1,v_2)\in\calE} b_{e}^{\text{ArbiTree}}\signfunc(\calU)_{v_1}\signfunc(\calU)_{v_2} \Big)
    \end{aligned}
\end{align} 
\begin{align}\label{eq:pgmscore}
    \begin{aligned} 
    s_{\text{PGM}}(\bX, \calU) &= - \Big( \sum_{v\in\calV} (c_{v}^{\text{PGM}})\T\varphi_v(\signfunc(\calU)_{v}, \bX) \\
    &+ \sum_{e = (v_1,v_2)\in\calE} (b_{e}^{\text{PGM}})\T\psi(\signfunc(\calU)_{v_1},\signfunc(\calU)_{v_2}). \Big)
    \end{aligned}
\end{align}

The ArbiTree score $s_{\text{ArbiTree}}$ assumes that the output of $f$ is directly a subset of $\calV$. The parameters $\set{c_{v}^{\text{ArbiTree}}\in\Real\given v\in\calV}\bigcup \set{b_{e}^{\text{ArbiTree}}\in\Real\given e\in\calE}$ are learnable. Both the tree structure and parameters are learned on a hold-out subset in the calibration set.

The PGM score $s_{\text{PGM}}$ assumes a function $\widetilde{f}_v$ on each $v\in\calV$, and then aggregates them to produce a non-conformity score for any subset $\calU\in\calV$. The learnable parameters are $\set{c_{v}^{\text{PGM}}\in\Real^2\given v\in\calV}\bigcup \set{b_{e}^{\text{PGM}}\in\Real^4\given e\in\calE}$. Similar to {\arbitr}, both the tree and the parameters are learned from the held-out calibration set.

Given these score designs and the calibration set, the prediction set is formed as follows \cite[Algorithm 2]{CauGupDuc:J21}:
\begin{align}\label{eq:cqioc}
    \widehat{C}_{\text{CQioC}}(\bX_{n+1};\alpha):=\set{v\given \min_{\calU:v\in\calU}s(\bX,\calU)\leq\widehat{q}_{\alpha}},
\end{align}
where
\begin{align*}
    \widehat{q}_{\alpha} := \mathrm{Q}(s(\bX_i,\calY_i), (1-\alpha)(1+\ofrac{n})).
\end{align*}
Here we have omitted the quantile regression step as in original {\cqioc}, or equivalently, we set the quantile regressor as a constant function that always has value $0$. The reasons are two-folds: first, the quantile regression is parallel with the prediction set formulation: given any score formulation, it is always possible to apply the quantile regression strategy to improve the performance of \gls{CP}. Second, our focus here is on prediction set formulation for the set inclusion problem. By investigating the pure prediction set formulation step, it can be understood in a more straightforward way. 

In general the set \cref{eq:cqioc} cannot be evaluated due to the high complexity of searching over all sets that contains a vertex $v$. However, given the tree structure and the tree-based scores $s_{\text{ArbiTree}}$ and $s_{\text{PGM}}$, this set can be efficiently calculated using message passing with complexity $O(N)$. For PGM, the procedure of learning the tree structure requires solving $O(N^2)$ optimization problems in total without closed-form solution, leading to an excessive computational load. Therefore in our experiment we only implement the {\arbitr} score.

\subsection{Difference and Connection}\label{subsect:diffconn}

We first explain the differences between this strategy and our proposed method. First, the above non-conformity score formulations do not operate through the function $\gamma$ (cf. \ \cref{eq:gam,eq:gam_based_score} in the main content). Second, the {\cqioc} prediction set \cref{eq:cqioc} is computationally prohibited without a tree structure. Therefore, both the score and prediction designs heavily rely on a tree structure. However, in complex networks, using trees may lead to an oversimplified model. This explains the inefficiency of {\arbitr} in the multi-source detection problem in general networks.

Next, we point out that when the non-conformity score fulfills our proposed design \cref{eq:gam_based_score,eq:mono_score}, the {\cqioc} formulation leads to the same prediction set as our formulation \cref{eq:confmset}. This is an immediate result by the following key observation:

\begin{Proposition}\label{prop:equi_cqioc}
    If the score function $s$ satisfies \cref{eq:gam_based_score,eq:mono_score}, then we have
    \begin{align*}
        \min_{\calU:v\in\calU}s(\bX,\calU) = s(\bX,\gamma(\bX,\set{v}))
    \end{align*}
    for any $v\in\calV$. In this case, {\cqioc} leads to the same prediction set as our proposed formulation \cref{eq:confmset}.
\end{Proposition}
\begin{proof}
    We first prove that $\min_{\calU:v\in\calU}s(\bX,\calU) \geq s(\bX,\gamma(\bX,\set{v}))$.     By the score design \cref{eq:gam_based_score} we know that $s(\bX,\calU) = s(\bX,\gamma(\bX,\calU))$, for any $\calU\subset\calV$.
    Since $v\in\calU$, we know that $\gamma(\bX, \set{v})\subset\gamma(\bX,\calU)$ by definition of $\gamma$. According to the monotonicity condition \cref{eq:mono_score}, we know that
    \begin{align*}
        s(\bX,\gamma(\bX,\set{v}))\leq s(\bX,\gamma(\bX,\calU)).
    \end{align*}
    Hence $\min_{\calU:v\in\calU}s(\bX,\calU) \geq s(\bX,\gamma(\bX,\set{v}))$.

    Next we prove that $\min_{\calU:v\in\calU}s(\bX,\calU) \leq s(\bX,\gamma(\bX,\set{v}))$, which is an immediate result by letting $\calU=\set{v}$. This completes the proof.
\end{proof}

This result indicates that, our score design is also able to reduce the computational complexity of \cref{eq:cqioc} to $O(N)$. More importantly, using our prposed score design, it is no longer necessary to assume a tree structure. 

\section{Comparison and Relation with \gls{CRC}}\label{sect:comp_crc}

In this section, we show that the proposed conformal set can be regarded as a score-based version of \gls{CRC}.

\subsection{Conformal set design}

Consider a parameterized prediction set $\calC_\lambda(\bX)$. Assume that 
    $\indicate*{\frac{\abs*{\calC_\lambda(\bX_i)\bigcap\calY_i}}{\abs*{\calY_i}} < \beta}$
is non-increasing in $\lambda$ for all $i\in[n]$.
It is proved that if $\widehat{\lambda}$ is chosen by \cite[Section 4.2]{AnaSteAda:C24}
\begin{align*}
    \widehat{\lambda} := \inf\set*{\lambda\given \ofrac{n+1}\sum_{i=1}^n \indicate*{\frac{\abs*{\calC_\lambda(\bX_i)\bigcap\calY_i}}{\abs*{\calY_i}} < \beta} + \ofrac{n+1} \leq \alpha},
\end{align*}
then the prediction set
\begin{align*}
    \widehat{C}_{\text{CRC}}(\bX_{n+1};\alpha,\beta) := \calC_{\widehat{\lambda}}(\bX_{n+1})
\end{align*}
satisfies
\begin{align*}
    \P(\frac{\abs*{\widehat{C}_{\text{CRC}}(\bX_{n+1};\alpha,\beta)\bigcap\calY_{n+1}}}{\abs*{\calY_{n+1}}}\geq 1-\beta) \geq 1-\alpha.
\end{align*}
\subsection{Equivalence and difference}

We first show that the proposed prediction set formulation \cref{eq:confmset_part} is a special construction of \gls{CRC}, with the parametric set family
\begin{align}\label{eq:score_ver_for_crc}
    \calC_\lambda(\bX) :
    = \set{v\given s(\bX, \nu(\bX, \set{v}))\leq \lambda}.
\end{align}
We utilize the following key observation:
\begin{Proposition}\label{prop:equiv_event}
    \begin{align*}
        \indicate*{\frac{\abs*{\calC_\lambda(\bX_i)\bigcap\calY_i}}{\abs*{\calY_i}} \geq 1- \beta} = \indicate*{\nu(\bX, \calY)\subset \calC_\lambda(\bX_i)}.
    \end{align*}
\end{Proposition}
\begin{proof}
    In the proof of \cref{thm:inclu_res_part} we have shown that $\nu(\bX, \calY)\subset \calC_\lambda(\bX_i)$ implies $\frac{\abs*{\calC_\lambda(\bX_i)\bigcap\calY_i}}{\abs*{\calY_i}} \geq 1- \beta$. We next show that the converse is also correct.

    As $\frac{\abs*{\calC_\lambda(\bX_i)\bigcap\calY_i}}{\abs*{\calY_i}} \geq 1- \beta$, we know that $\calC_\lambda(\bX_i)$ has already included at least $1- \beta$ proportion of $\calY$. So it suffices to argue that $\calC_\lambda(\bX_i)$ includes those $1- \beta$ proportion of $\calY$ with the highest predicted probabilities, which is $\nu(\bX, \calY)$. This can be proved by contradiction. Suppose there exists a vertex $v'$ so that $v'\in\nu(\bX, \calY)$ but $v'\notin \calC_\lambda(\bX)$. By definition of $\calC_\lambda(\bX)$, we know that $s(\bX, \nu(\bX, \set{v'}))> -\lambda$. For any $v''\in\calV$ with $\widehat{\pi}(v'')\leq \widehat{\pi}(v')$, by definition of $\nu$ (cf. \ \cref{eq:ordered_shrink}), we know that $\nu(\bX, \set{v'})\subset \nu(\bX, \set{v''})$. By monotonicity of the non-conformity score (cf. \ \cref{eq:mono_score}), we know that $s(\bX, \nu(\bX, \set{v'}))\leq s(\bX, \nu(\bX, \set{v''}))$. Therefore $s(\bX, \nu(\bX, \set{v''}))> -\lambda$, thus $v''\notin \calC_\lambda(\bX)$. As $v'$ belongs to the top $1-\beta$ true sources with large predicted probability, $\calC_\lambda(\bX)$ can only contain nodes with larger predicted probability than $v'$, which is strictly less than $1-\beta$ proportion of $\calY$, leading to contradiction.
\end{proof}

Applying \cref{prop:equiv_event} we see that 
\begin{align*}
    \widehat{\lambda} &= \inf\set*{\lambda\given \ofrac{n}\sum_{i=1}^n\indicate*{\frac{\abs*{\calC_\lambda(\bX_i)\bigcap\calY_i}}{\abs*{\calY_i}} < \beta}\leq (1+\ofrac{n})\alpha - \ofrac{n}}\\
    &= \inf\set*{\lambda\given \ofrac{n}\sum_{i=1}^n\indicate*{\frac{\abs*{\calC_\lambda(\bX_i)\bigcap\calY_i}}{\abs*{\calY_i}} \geq 1- \beta}\geq (1+\ofrac{n})(1-\alpha)}\\
    &=\inf\set*{\lambda\given \ofrac{n}\sum_{i=1}^n\indicate*{\nu(\bX_i, \calY_i)\subset \calC_\lambda(\bX_i)}\geq (1+\ofrac{n})(1-\alpha)} \\
    &= \inf\set*{\lambda\given \ofrac{n}\sum_{i=1}^n\indicate*{s(\bX_i, \nu(\bX_i, \calY_i))\leq \lambda}\geq (1+\ofrac{n})(1-\alpha)} \\
    &= \mathrm{Q}(s(\bX_i,\calY_i), (1-\alpha)(1+\ofrac{n})) \\
    &= \widehat{q}_\alpha.
\end{align*}
Combining this result with \cref{eq:score_ver_for_crc} we obtain the same prediction set with \cref{eq:confmset_part}.

Next, we show that \gls{CRC} is a specific version of \cref{eq:confmset_part} with the following specific non-conformity score design:
\begin{align}\label{eq:score_for_crc}
    s(\bX, \calU) := \inf\set*{\lambda\given \calU\subset \calC_\lambda(\bX)}
\end{align}
We assume the following mild conditions on $\calC_\lambda(\bX)$:
\begin{Assumption}\label{asp:para_set_family}\
    \begin{enumerate}[(i)]
    \item \label[condition]{cond:monot_set} $\calC_\lambda(\bX)$ is monotone \gls{wrt} $\lambda$, i.e., $\calC_{\lambda_1}(\bX)\leq \calC_{\lambda_2}(\bX)$ for any $\lambda_1\leq \lambda_2$.
    \item \label[condition]{cond:monot_prob} For any $\calU\subset\calV$, $\calU\subset \calC_\lambda(\bX)$ implies $\gamma(\calU)\subset\calC_\lambda(\bX)$.
    \item \label[condition]{cond:monot_para} $\abs*{\calC_\lambda(\bX)}$ is right-continuous on $\lambda$.
\end{enumerate}
\end{Assumption}
\Cref{cond:monot_set} requires the parametric set family to be nested. Note that this is a sufficient condition for $\indicate*{\frac{\abs*{\calC_\lambda(\bX_i)\bigcap\calY_i}}{\abs*{\calY_i}} < \beta}$
being non-increasing in $\lambda$ for all $i\in[n]$. \Cref{cond:monot_prob} requires the parametric set family to be those sets of vertices with top largest predicted probabilities. \Cref{cond:monot_para} requires continuity of the set \gls{wrt} $\lambda$. One important example is 
\begin{align}\label{eq:crc_set}
    \calC_\lambda(\bX) := \set*{v\in\calV\given \widehat{\pi}(v)\geq 1-\lambda}.
\end{align}
This set construction is proposed for controlling the recall rate in \cite{AnaSteAda:C24}. We have the following key observation similar as \cref{prop:equiv_event}:

\begin{Proposition}\label{prop:equi_score_and_event}
    Under \cref{asp:para_set_family}, we have
    \begin{align}
            s(\bX, \calU) &= \min\set*{\lambda\given \calU\subset \calC_\lambda(\bX)} \label{eq:inf_to_min}\\ 
        \indicate*{\frac{\abs*{\calC_\lambda(\bX)\bigcap\calY}}{\abs*{\calY}} \geq 1- \beta} &= \indicate*{\nu(\bX, \calY)\subset \calC_\lambda(\bX)}. \label{eq:identical_event}
    \end{align}
\end{Proposition}
\begin{proof}
    We first prove \cref{eq:inf_to_min} by showing that $\calU\subset \calC_{s(\bX,\calU)}(\bX)$. Since $\abs*{\calC_\lambda(\bX)}\in\bbN$ and $\abs*{\calC_\lambda(\bX)}$ is right-continuous and non-decreasing in $\lambda$, we know that there exists $\delta > 0$ such that $\abs*{\calC_\lambda(\bX)} = \abs*{\calC_{s(\bX,\calU)}(\bX)}$ for all $\lambda\in(s(\bX, \calU), s(\bX, \calU) + \delta)$. Recall that $\calC_\lambda(\bX)$ is non-decreasing in $\lambda$, hence $\calC_\lambda(\bX) = \calC_{s(\bX,\calU)}(\bX)$ for all $\lambda\in(s(\bX, \calU), s(\bX, \calU) + \delta)$. As $\calU\in \calC_\lambda(\bX)$ for all $\lambda\in(s(\bX, \calU), s(\bX, \calU) + \delta)$, we know that $\calU\subset \calC_{s(\bX,\calU)}(\bX)$. By definition \cref{eq:score_for_crc} we obtain \cref{eq:inf_to_min}.

    The proof of \cref{eq:identical_event} is similar with that of \cref{prop:equiv_event}. In the proof of \cref{thm:inclu_res_part} we have shown that $\nu(\bX, \calY)\subset \calC_\lambda(\bX)$ implies $\frac{\abs*{\calC_\lambda(\bX)\bigcap\calY}}{\abs*{\calY}} \geq 1- \beta$. The converse direction can be proved by directly applying \cref{cond:monot_prob} in \cref{asp:para_set_family}.
\end{proof}

Applying \cref{prop:equi_score_and_event} we have 
\begin{align*}
    \widehat{q}_\alpha &= \mathrm{Q}(s(\bX_i,\nu(\bX, \calY_i)), (1-\alpha)(1+\ofrac{n})) \\
    &= \inf\set*{\lambda\given \ofrac{n}\sum_{i=1}^n \indicate*{s(\bX_i,\nu(\bX, \calY_i))\leq \lambda} \geq (1-\alpha)(1+\ofrac{n})} \\
    &= \inf\set*{\lambda\given \ofrac{n}\sum_{i=1}^n \indicate*{\nu(\bX, \calY_i)\subset\calC_\lambda(\bX_i)} \geq (1-\alpha)(1+\ofrac{n})} \\
    &= \inf\set*{\lambda\given \ofrac{n}\sum_{i=1}^n \indicate*{\frac{\abs*{\calC_\lambda(\bX_i)\bigcap\calY_i}}{\abs*{\calY_i}} \geq 1- \beta} \geq (1-\alpha)(1+\ofrac{n})} \\
    &= \widehat{\lambda}.
\end{align*}

Combining this result with \cref{prop:equi_cqioc}, we know that under \cref{asp:para_set_family}, \gls{CRC} also leads to the same prediction set with {\cqioc} using out proposed equivalent score design principles \cref{eq:mono_score}.

Finally, we show that the prediction set \cref{eq:crc_set} corresponds to using the non-conformity score \cref{eq:min_score} in our proposed prediction set \cref{eq:confmset_part}. Using \cref{eq:inf_to_min} we know that 
\begin{align*}
    s(\bX, \calU) = 1 - \min_{u\in\calU}\widehat{\pi}(\calU) = 1 - \min_{u\in\gamma(\bX, \calU)}\widehat{\pi}(\calU),
\end{align*}
which differs with \cref{eq:min_score} by only a constant. Therefore, the prediction set \cref{eq:crc_set} proposed by \cite{AnaSteAda:C24} for controlling the recall rate is the same with our proposed framework with the score design \cref{eq:min_score} .

\section{Experimental Settings}\label{sect:expset}

\subsection{Simulation Configurations}

In this subsection, we describe the simulator configurations. We first introduce the graphs highSchool, bkFratB, and sfhh. They are respectively contact social networks between people in an American highschool, students living in a fraternity at a West Virginia college, and participants at a conference. Their statistics are summarized in \cref{tab:graphs} \cite{ShaHasMoh:C21}:

\begin{table}[htbp]
    \centering
    \small
    \caption{Graph statistics. The columns show number of nodes $\abs{\calV}$, number of edges $\abs{\calE}$ and average degree.}
    \setlength{\tabcolsep}{2pt} 
    \begin{tabular}{p{1.8cm} p{1.8cm} p{2.3cm} p{1.95cm}}
    \toprule
     & $\abs{\calV}$ & $\abs{\calE}$ & average degree \\
     {highSchool} & 774 & 7992 & 20.7 \\
     {bkFratB} & 58 & 967 & 33.3 \\
     {sfhh} & 403 & 9565 & 47.5 \\
    \bottomrule
    \end{tabular}
    
    \label{tab:graphs}
\end{table}

We next describe the SIR simulator used in the experiments. The propagation processes are simulated via NDlib,\footnote{\url{https://ndlib.readthedocs.io/en/latest/index.html}} a package for simulating diffusion processes over networks. When the number of sources is given, the source nodes are uniformly randomly drawn from all nodes. Then the simulator generates propagation paths with the parameters $\infectrate$ and $\recoverrate$ for consecutive discrete time steps. At the $0$-th time instance, all sources have status I, except for those source nodes that have status S. The observation begins at some instance $t_1>0$, and lasts for $16$ consecutive time instances. In \cref{tab:nomilevels,tab:si_sir,tab:diffgraphs} of the main content we have described the configurations of number of sources, $\infectrate$ and $\recoverrate$ (can be computed from $R_0$). Here, we specify the starting instance of observation, i.e., $t_1>0$. In this experiment, when the number of sources equals $1$ or $R_0\in[1, 15]$, we set $t_1=2$, otherwise $t_1 = 1$. By this setting we take into account the fact that a small number of sources and slow propagation may lead to later awareness of the epidemic. 

\subsection{Implementation Details}

The experiments are conducted on a Ubuntu 18.04.4 LTS (Bionic Beaver) workstation with a single NVIDIA RTX 2080Ti GPU ($11$ GB memory) and Intel Xeon W-2133 CPU with $12$ cores.
The pre-trained graph neural network is implemented using the TensorFlow 2.3.0 framework.

The {\adit} method is implemented by the source code \footnote{https://github.com/lab-sigma/Diffusion-Source-Identification}. The hyperparameters are $m_l$ and $m_p$. $m_l$ adjusts the number of simulation paths originating from each node, and $m_p$ adjusts the number of Monte Carlo runs for estimating the empirical distribution of the test statistics. In this experiment, to achieve balance between execution time and prediction quality, we choose $m_l = 20$, $m_p = 20$.

The {\arbitr} method requires a hold-out set for learning the tree structure. In this experiment we set the size of this hold-out set to be $1000$. We note that, the number of parameters need to be estimated equals $\ofrac{2}\abs{\calV}(\abs{\calV}+1)$, this estimation problem is easily underdetermined when the graph is large.

\section{Additional Experimental Results}\label{sect:addexpres}

\begin{table*}[htbp]
    \centering
    \small 
    \caption{Inclusion rates and prediction set sizes under the SIR model with random parameters ($|\mathcal{Y}| \in [15]$, $R_0 \in [11, 25]$, $\recoverrate \in [0.1, 0.4]$) over the highSchool network for methods \setprec, \setrec, and \arbitr. The best and the second-best results under each $(\alpha,\beta)$ configuration are highlighted by bold faces and underlined respectively.} 
    \begin{tabular}{ll *{6}{c@{\hspace{0.3cm}}}}
        \toprule
        & & \multicolumn{3}{c}{\textbf{Inclusion Rates}} & \multicolumn{3}{c}{\textbf{Prediction Set Size}} \\
        \cmidrule(lr){3-5} \cmidrule(lr){6-8}
        \textbf{Method} & \textbf{$\beta$} & \textbf{$\alpha=0.05$} & \textbf{$\alpha=0.10$} & \textbf{$\alpha=0.15$} & \textbf{$\alpha=0.05$} & \textbf{$\alpha=0.10$} & \textbf{$\alpha=0.15$} \\
        \midrule
        \multirow{4}{*}{\textbf{\setrec}} & $0.1$ & 1.000 $\pm$ 0.000 & 0.899 $\pm$ 0.015 & 0.849 $\pm$ 0.021 & 774.000 $\pm$ 0.000 & \underline{33.461 $\pm$ 1.016} & \underline{32.424 $\pm$ 0.979} \\
        & $0.3$ & 0.947 $\pm$ 0.011 & 0.894 $\pm$ 0.018 & 0.845 $\pm$ 0.022 & \underline{30.038 $\pm$ 0.909} & \underline{26.340 $\pm$ 0.783} & \underline{24.397 $\pm$ 0.728} \\
        & $0.5$ & 0.949 $\pm$ 0.009 & 0.899 $\pm$ 0.012 & 0.849 $\pm$ 0.017 & \textbf{18.951 $\pm$ 0.555} & \textbf{15.558 $\pm$ 0.460} & \textbf{13.793 $\pm$ 0.390} \\
        & $0.7$ & 0.950 $\pm$ 0.009 & 0.900 $\pm$ 0.015 & 0.850 $\pm$ 0.019 & \textbf{14.207 $\pm$ 0.412} & \textbf{9.663 $\pm$ 0.282} & \textbf{7.509 $\pm$ 0.215} \\
        \midrule
        \multirow{4}{*}{\textbf{\setprec}} & $0.1$ & 0.950 $\pm$ 0.011 & 0.900 $\pm$ 0.016 & 0.847 $\pm$ 0.020 & \underline{43.925 $\pm$ 1.034} & 38.953 $\pm$ 0.975 & 35.512 $\pm$ 0.934 \\
        & $0.3$ & 0.950 $\pm$ 0.011 & 0.898 $\pm$ 0.018 & 0.846 $\pm$ 0.020 & 37.554 $\pm$ 0.962 & 31.705 $\pm$ 0.902 & 28.275 $\pm$ 0.852 \\
        & $0.5$ & 0.948 $\pm$ 0.012 & 0.897 $\pm$ 0.017 & 0.845 $\pm$ 0.019 & 27.865 $\pm$ 0.869 & 22.367 $\pm$ 0.768 & 18.680 $\pm$ 0.706 \\
        & $0.7$ & 0.949 $\pm$ 0.011 & 0.898 $\pm$ 0.015 & 0.848 $\pm$ 0.018 & 21.629 $\pm$ 0.766 & 15.403 $\pm$ 0.632 & 11.649 $\pm$ 0.515 \\
        \midrule
        \multirow{4}{*}{\textbf{\setmin}} & $0.1$ &  0.949 ± 0.012  &  0.899 ± 0.015  &  0.850 ± 0.019  & \textbf{33.283 ± 0.858} & \textbf{31.459 ± 0.758} & \textbf{29.739 ± 0.703}  \\
        & $0.3$ &  0.951 ± 0.010  &  0.901 ± 0.015  &  0.847 ± 0.019  & \textbf{28.982 ± 0.697} & \textbf{26.091 ± 0.654} & \textbf{23.841 ± 0.661} \\
        & $0.5$ &  0.951 ± 0.011  &  0.896 ± 0.015  &  0.846 ± 0.020  & \underline{22.273 ± 0.631} & \underline{18.490 ± 0.603} & \underline{15.959 ± 0.593} \\
        & $0.7$ &  0.949 ± 0.012  &  0.898 ± 0.016  &  0.846 ± 0.020  & \underline{15.676 ± 0.606} & \underline{11.640 ± 0.536} & \underline{9.233 ± 0.468} \\
        \midrule
        \multirow{4}{*}{\textbf{\arbitr}} & $0.1$ & 0.995 $\pm$ 0.003 & 0.991 $\pm$ 0.006 & 0.982 $\pm$ 0.009 & 772.191 $\pm$ 0.565 & 769.953 $\pm$ 1.139 & 766.208 $\pm$ 2.312 \\
        & $0.3$ & 0.998 $\pm$ 0.002 & 0.992 $\pm$ 0.004 & 0.977 $\pm$ 0.007 & 769.957 $\pm$ 1.763 & 761.811 $\pm$ 3.511 & 746.261 $\pm$ 6.768 \\
        & $0.5$ & 0.992 $\pm$ 0.004 & 0.971 $\pm$ 0.010 & 0.935 $\pm$ 0.015 & 759.565 $\pm$ 4.205 & 733.803 $\pm$ 10.268 & 698.095 $\pm$ 13.750 \\
        & $0.7$ & 0.977 $\pm$ 0.008 & 0.937 $\pm$ 0.013 & 0.892 $\pm$ 0.019 & 734.878 $\pm$ 8.993 & 691.744 $\pm$ 14.236 & 645.393 $\pm$ 14.612 \\
        \bottomrule
    \end{tabular}
    
    \label{tab:nomilevels_mid}
\end{table*}

\begin{table*}[htbp]
    \centering
    \small 
    \caption{Inclusion rates and prediction set sizes under the SIR model with random parameters ($|\mathcal{Y}| \in [15]$, $R_0 \in [21, 35]$, $\recoverrate \in [0.1, 0.4]$) over the highSchool network for methods \setprec, \setrec, and \arbitr. The best and the second-best results under each $(\alpha,\beta)$ configuration are highlighted by bold faces and underlined respectively.} 
    \begin{tabular}{ll *{6}{c@{\hspace{0.3cm}}}}
        \toprule
        & & \multicolumn{3}{c}{\textbf{Inclusion Rates}} & \multicolumn{3}{c}{\textbf{Prediction Set Size}} \\
        \cmidrule(lr){3-5} \cmidrule(lr){6-8}
        \textbf{Method} & \textbf{$\beta$} & \textbf{$\alpha=0.05$} & \textbf{$\alpha=0.10$} & \textbf{$\alpha=0.15$} & \textbf{$\alpha=0.05$} & \textbf{$\alpha=0.10$} & \textbf{$\alpha=0.15$} \\
        \midrule
        \multirow{4}{*}{\textbf{\setrec}} & $0.1$ & 0.975 $\pm$ 0.030 & 0.898 $\pm$ 0.013 & 0.848 $\pm$ 0.014 & 483.524 $\pm$ 355.760 & \underline{46.514 $\pm$ 1.588} & \underline{45.125 $\pm$ 1.544} \\
        & $0.3$ & 0.947 $\pm$ 0.010 & 0.897 $\pm$ 0.015 & 0.846 $\pm$ 0.020 & \textbf{41.402 $\pm$ 1.438} & \textbf{37.055 $\pm$ 1.270} & \textbf{34.002 $\pm$ 1.161} \\
        & $0.5$ & 0.950 $\pm$ 0.010 & 0.899 $\pm$ 0.012 & 0.848 $\pm$ 0.017 & \textbf{27.088 $\pm$ 0.930} & \textbf{21.668 $\pm$ 0.728} & \textbf{18.706 $\pm$ 0.633} \\
        & $0.7$ & 0.952 $\pm$ 0.010 & 0.901 $\pm$ 0.016 & 0.852 $\pm$ 0.020 & \textbf{19.429 $\pm$ 0.688} & \textbf{12.829 $\pm$ 0.461} & \textbf{9.845 $\pm$ 0.348} \\
        \midrule
        \multirow{4}{*}{\textbf{\setprec}} & $0.1$ & 0.950 $\pm$ 0.011 & 0.900 $\pm$ 0.015 & 0.851 $\pm$ 0.015 & \underline{63.138 $\pm$ 1.958} & 57.418 $\pm$ 1.795 & 53.349 $\pm$ 1.697 \\
        & $0.3$ & 0.952 $\pm$ 0.012 & 0.904 $\pm$ 0.016 & 0.853 $\pm$ 0.019 & 55.449 $\pm$ 1.751 & 47.044 $\pm$ 1.553 & 41.446 $\pm$ 1.404 \\
        & $0.5$ & 0.952 $\pm$ 0.011 & 0.903 $\pm$ 0.014 & 0.852 $\pm$ 0.019 & 42.047 $\pm$ 1.444 & 32.006 $\pm$ 1.134 & 26.426 $\pm$ 0.963 \\
        & $0.7$ & 0.952 $\pm$ 0.011 & 0.902 $\pm$ 0.015 & 0.849 $\pm$ 0.018 & 31.059 $\pm$ 1.133 & 20.739 $\pm$ 0.829 & 14.972 $\pm$ 0.626 \\
        \midrule
        \multirow{4}{*}{\textbf{\setmin}} & $0.1$ & 0.949 ± 0.011 & 0.899 ± 0.016 & 0.849 ± 0.019 & \textbf{47.506 ± 1.589}  & \textbf{45.491 ± 1.506} & \textbf{43.652 ± 1.441} \\
        & $0.3$ & 0.952 ± 0.013 & 0.903 ± 0.014 & 0.854 ± 0.017 & \underline{41.733 ± 1.386} & \underline{37.698 ± 1.262} & \underline{34.743 ± 1.203} \\
        & $0.5$ & 0.953 ± 0.010 & 0.904 ± 0.015 & 0.854 ± 0.016 & \underline{31.172 ± 1.120} & \underline{25.722 ± 0.972} & \underline{22.239 ± 0.876} \\
        & $0.7$ & 0.952 ± 0.011 & 0.903 ± 0.016 & 0.853 ± 0.020 & \underline{21.682 ± 0.879} & \underline{15.806 ± 0.711} & \underline{11.895 ± 0.601} \\
        \midrule
        \multirow{4}{*}{\textbf{\arbitr}} & $0.1$ & 0.997 $\pm$ 0.002 & 0.993 $\pm$ 0.003 & 0.987 $\pm$ 0.005 & 772.491 $\pm$ 0.284 & 770.773 $\pm$ 1.007 & 767.010 $\pm$ 2.488 \\
        & $0.3$ & 0.998 $\pm$ 0.002 & 0.990 $\pm$ 0.006 & 0.973 $\pm$ 0.009 & 770.857 $\pm$ 1.275 & 762.320 $\pm$ 4.470 & 746.111 $\pm$ 7.571 \\
        & $0.5$ & 0.990 $\pm$ 0.005 & 0.963 $\pm$ 0.011 & 0.934 $\pm$ 0.015 & 760.536 $\pm$ 4.572 & 733.977 $\pm$ 9.625 & 706.730 $\pm$ 12.725 \\
        & $0.7$ & 0.972 $\pm$ 0.009 & 0.932 $\pm$ 0.015 & 0.891 $\pm$ 0.017 & 738.028 $\pm$ 9.134 & 698.984 $\pm$ 13.043 & 659.134 $\pm$ 14.804 \\
        \bottomrule
    \end{tabular}
    
    \label{tab:nomilevels_Fast}
\end{table*}

\begin{table*}[htbp]
    \centering
    \small 
    \caption{Inclusion rates and prediction set sizes under the SI model ($\infectrate=0.25$) and SIR model ($\infectrate=0.25$, $\recoverrate=0.15$) over the highSchool network with different number of sources (denoted by $\abs{\calY}$). Nominal levels are set as $\alpha=0.1$, $\beta=0.1$.}
    \begin{tabular}{ll *{4}{c@{\hspace{0.3cm}}}}
        \toprule
        & & \multicolumn{2}{c}{\textbf{Inclusion Rates}} & \multicolumn{2}{c}{\textbf{Prediction Set Size}} \\
        \cmidrule(lr){3-4} \cmidrule(lr){5-6}
        \textbf{Model} & \textbf{Method} & \textbf{$|\mathcal{Y}|=7$} & \textbf{$|\mathcal{Y}|=10$} & \textbf{$|\mathcal{Y}|=7$} & \textbf{$|\mathcal{Y}|=10$} \\
        \midrule
        \multirow{3}{*}{\textbf{SI}} 
        & \setrec & 0.901 $\pm$ 0.015 & 0.902 $\pm$ 0.015 & \underline{40.424 $\pm$ 0.357} & \textbf{52.401 $\pm$ 0.511} \\
        & \setprec & 0.901 $\pm$ 0.015 & 0.904 $\pm$ 0.014 & 42.875 $\pm$ 0.264 & 57.716 $\pm$ 0.450 \\
        & \setmin & 0.904 ± 0.016 & 0.902 ± 0.014 & \textbf{40.285 ± 0.355} & \underline{52.412 ± 0.596} \\
        & \arbitr & 0.969 $\pm$ 0.021 & 0.997 $\pm$ 0.004 & 770.128 $\pm$ 2.346 & 769.069 $\pm$ 2.518 \\
        \midrule
        \multirow{4}{*}{\textbf{SIR}} & \textbf{Method} & $|\mathcal{Y}|\!=\!7$ & $|\mathcal{Y}|\!=\!10$ & $|\mathcal{Y}|\!=\!7$ & $|\mathcal{Y}|\!=\!10$ \\
        \cmidrule(lr){2-6}
        & \setrec & 0.902 $\pm$ 0.016 & 0.901 $\pm$ 0.016 & \underline{40.300 $\pm$ 0.381} & \textbf{51.463 $\pm$ 0.379} \\
        & \setprec & 0.904 $\pm$ 0.018 & 0.898 $\pm$ 0.016 & 42.516 $\pm$ 0.263 & 57.950 $\pm$ 0.368 \\
        & \setmin & 0.903 ± 0.016 & 0.900 ± 0.015 & \textbf{39.952 ± 0.425} & \underline{51.918 ± 0.464} \\
        & \arbitr & 0.964 $\pm$ 0.019 & 0.997 $\pm$ 0.003 & 769.432 $\pm$ 2.257 & 768.101 $\pm$ 2.773 \\
        \bottomrule
    \end{tabular}
    
    \label{tab:si_sir_pow9}
\end{table*}

\begin{table*}[htbp]
    \centering
    \small 
    \caption{Inclusion rates and prediction set sizes under the SI model ($\infectrate=0.25$) and SIR model ($\infectrate=0.25$, $\recoverrate=0.15$) over the highSchool network with different number of sources (denoted by $\abs{\calY}$). Nominal levels are set as $\alpha=0.1$, $\beta=0.5$.}
    \begin{tabular}{ll *{4}{c@{\hspace{0.3cm}}}}
        \toprule
        & & \multicolumn{2}{c}{\textbf{Inclusion Rates}} & \multicolumn{2}{c}{\textbf{Prediction Set Size}} \\
        \cmidrule(lr){3-4} \cmidrule(lr){5-6}
        \textbf{Model} & \textbf{Method} & \textbf{$|\mathcal{Y}|=7$} & \textbf{$|\mathcal{Y}|=10$} & \textbf{$|\mathcal{Y}|=7$} & \textbf{$|\mathcal{Y}|=10$} \\
        \midrule
        \multirow{3}{*}{\textbf{SI}} 
        & \setrec & 0.903 $\pm$ 0.015 & 0.900 $\pm$ 0.014 & \textbf{22.778 $\pm$ 0.247} & \textbf{27.764 $\pm$ 0.336} \\
        & \setprec & 0.901 $\pm$ 0.015 & 0.902 $\pm$ 0.016 & 34.142 $\pm$ 0.428 & 44.695 $\pm$ 0.619 \\
        & \setmin & 0.900 ± 0.014 & 0.902 ± 0.014 & \underline{25.710 ± 0.254} & \underline{32.668 ± 0.279} \\
        & \arbitr & 0.999 $\pm$ 0.001 & 0.998 $\pm$ 0.003 & 767.366 $\pm$ 1.808 & 764.446 $\pm$ 3.518 \\
        \midrule
        \multirow{4}{*}{\textbf{SIR}} & \textbf{Method} & $|\mathcal{Y}|\!=\!7$ & $|\mathcal{Y}|\!=\!10$ & $|\mathcal{Y}|\!=\!7$ & $|\mathcal{Y}|\!=\!10$ \\
        \cmidrule(lr){2-6}
        & \setrec & 0.901 $\pm$ 0.016 & 0.900 $\pm$ 0.013 & \textbf{21.351 $\pm$ 0.253} & \textbf{23.927 $\pm$ 0.223} \\
        & \setprec & 0.903 $\pm$ 0.013 & 0.899 $\pm$ 0.015 & 29.228 $\pm$ 0.313 & 37.363 $\pm$ 0.590 \\
        & \setmin & 0.902 ± 0.016 & 0.900 ± 0.015 & \underline{23.132 ± 0.318} & \underline{30.031 ± 0.461} \\
        & \arbitr & 1.000 $\pm$ 0.001 & 1.000 $\pm$ 0.000 & 764.505 $\pm$ 2.192 & 765.992 $\pm$ 1.241 \\
        \bottomrule
    \end{tabular}
    
    \label{tab:si_sir_pow5}
\end{table*}

In \cref{tab:nomilevels_mid,tab:nomilevels_Fast} we present the numerical results for different combinations of $(\alpha, \beta)$ under different ranges of $R_0$. The results lead to the same conclusion as in \cref{subsect:expnominals}. \Cref{tab:si_sir_pow9,tab:si_sir_pow5} show the numerical results under different nominal levels of $\beta$. This complements the results in \cref{subsect:exppropmods}, further validating the conclusion.

\bibliographystyle{IEEEtran}
\bibliography{refs,StringDefinitions,IEEEabrv}

\end{document}